\documentclass[12pt]{article}
\usepackage{pdproc} 

\begin{document}

\renewcommand{\thefootnote}{\alph{footnote}}

\title{ SEARCHING FOR DARK MATTER }

\author{ M. RONCADELLI}

\address{INFN, Pavia, Italy\\ 
{\rm E-mail: roncadelli@pv.infn.it}}

\abstract{The observational evidence for dark matter on progressively larger 
cosmic scales is reviewed in a rather pedagogical fashion. Although the 
emphasis is on dark matter in galaxies and in clusters of galaxies, its 
cosmological evidence as well as its physical nature are also discussed.}

\normalsize\baselineskip=15pt

\section{Introduction}

Almost all of our information about the Universe comes from electromagnetic 
radiation 
at different wavelenghts. Because we obviously miss a lot of photons, the 
existence of {\it dark} matter should hardly be surprising. Clearly, the only way 
to infer its presence rests upon the resulting gravitational effects on luminous 
matter. This strategy has a long tradition in astronomy. For instance, in 1846 it 
led to the discovery of Neptune from unexplained residuals in the motion of Urans. 
In a similar way, in 1933 Zwicky pointed out that the very existence of the Coma 
cluster of galaxies would be impossible unless its dynamics were dominated by dark 
matter. Regretfully, it took nearly four decades before Zwicky's suggestion became 
a respectable research topic. Today, we know that dark matter largely 
outweights luminous mass. Besides, we know that most of the dark matter differs 
drastically from ordinary stuff. Ironically, while still waiting to know what dark 
matter really is, we know for sure that it is responsible for the formation of all 
structure in the Universe, and so ultimately for our existence as well.

This talk describes in a rather pedagogical manner the observational evidence for 
dark matter on progressively larger cosmic scales. Emphasis is given to dark 
matter in galaxies and in clusters of galaxies. The cosmological evidence and the 
physical nature of dark matter are discussed more briefly (these topics are 
addressed also by other speakers at this conference). Throughout, the conservative 
attitude is taken that the Universe is described by known physical 
laws. Considerably more attention is paid to basic physical principles than to 
observing techniques. In order to keep the number of references under control, we 
largely quote review papers, from which the original references can easily be traced.

\section{Cosmological perspective}

According to the hot big bang cosmology~\cite{kn:PEACOCK}, the Universe is 
described by the 
Friedmann-Robertson-Walker model, which emerges from general relativity through the 
Cosmological Principle. In terms of a set of comoving coordinates, the space-time 
line element can be written as
\begin{equation}
ds^2 = c^2 dt^2 - R^2(t) \left[ \frac{dr^2}{1 - k  r^2} + r^2 \Bigl( d \theta^2 
+ \sin^2 \theta \, d \phi^2 \Bigr) \right]~,   \label{cos13}   
\end{equation}
where $R(t)$ is the cosmic scale factor and the constant $k$ is proportional to 
the gaussian curvature of tridimensional space. Then it follows that 
tridimensional space is {\it open} for $k < 0$, {\it flat} ({\it euclidean}) for 
$k = 0$ and {\it closed} for $k > 0$.

The dynamics of the Universe is parametrized by the cosmic scale factor, which 
obeys the equations that arise by inserting the metric dictated by eq. 
(\ref{cos13}) into Einstein equations, namely
\begin{equation}
\left( \frac{\dot R}{R} \right)^2 = \frac{8 \pi G}{3} \rho  - \frac{k  c^2}
{R^2}~,   \label{co22}   
\end{equation}
\begin{equation}
\frac{\ddot R}{R} = - \frac{4 \pi G}{3} \left(\rho + \frac{3 p}{c^2}\right)~,   
\label{co18}   
\end{equation}
where $\rho$ and $p$ denote the cosmic mass density~\footnote{More generally, 
$\rho$ is the cosmic energy density divided by $c^2$.} and pressure (respectively). 
Observe that by combining eqs. (\ref{co22}) and (\ref{co18}) together, the 
following conservation equation arises
\begin{equation}
\frac{d}{dt} \left( \rho  R^3 \right) = - \frac{3 p}{c^2} R^2 \dot R~.   
\label{co4b}   
\end{equation}
What remains to be specified at this point is the equation of state of the 
Universe, namely the relationship between $\rho$ and $p$. It is useful to write 
such a relation in the form
\begin{equation}
p = w \, \rho \, c^2~,   \label{CCco4b}   
\end{equation}
where $w$ -- which can be supposed {\it constant} -- takes the values $0$ for 
nonrelativistic {\it matter}, $1/3$ for {\it radiation} and $- 1$ in the presence 
of a nonvanishing vacuum energy described by a {\it cosmological constant}. 
Notice that in the latter case, $\rho$ and $p$ have {\it opposite} signs. 
By inserting eq. (\ref{CCco4b}) into eq. (\ref{co4b}), we get 
\begin{equation}
\rho \sim R^{ -3(1 + w)}~.   \label{co4c}   
\end{equation}
Hence, $\rho \sim R^{- 3}$ for nonrelativistic matter, $\rho \sim R^{- 4}$ for 
radiation and $\rho$ = constant for a cosmological constant. 

Now, it is very convenient to define the {\it Hubble parameter}
\begin{equation}
H \equiv \frac{\dot R}{R}~,   \label{co4c}   
\end{equation}
the {\it critical density}
\begin{equation}
{\rho}_c \equiv \frac{3  H^2}{8 \pi G}~,   \label{Uco4c}   
\end{equation}
and the {\it cosmic density parameter}
\begin{equation}
\Omega \equiv \frac{\rho}{\rho_c}~,   \label{Zco4c}   
\end{equation}
in terms of which eq. (\ref{co22}) can be rewritten as
\begin{equation}
\Omega = 1 + \frac{k c^2}{H^2 R^2}~.  \label{ZZco4c}   
\end{equation}
Consequently, the knowledge of $\Omega$ at a particular cosmic time -- which 
may be taken to be just the present -- is crucial to determine the {\it geometry} 
of the Universe. Indeed, recalling the connection between $k$ and geometry, we 
find that the Universe is {\it open} for $\Omega < 1$, {\it flat} ({\it 
euclidean}) for $\Omega = 1$ and  {\it closed} for $\Omega > 1$.

So far, no assumption has been made about $\rho$ and $p$. Suppose now that they 
are both {\it positive}, meaning that the Universe only contains {\it ordinary} 
matter and radiation (a cosmological constant is accordingly ruled out). Then the 
knowledge of $\Omega$ {\it also} determines the {\it evolution} of the Universe 
{\it uniquely}. For, in such a situation $\rho$ decreases faster than $R^{- 2}$ as $R$ 
increases. So, by eq. (\ref{co22}) $\dot R$ never vanishes as long as $k \leq 0$, 
whereas it does vanish at some cosmic time for $k > 0$. Therefore, an open or flat 
Universe ($\Omega \leq 1$) {\it expands forever}, while a closed Universe 
($\Omega > 1$) eventually {\it recollapses}. Put in a slightly different 
fashion, {\it geometry is destiny} as long as the cosmological constant vanishes. 
Another crucial 
feature of a Universe without cosmological constant is that the evolution is 
necessarily {\it decelerated}. This immediately follows from eq. (\ref{co18}) and 
can be traced to the fact that ordinary gravity is always {\it attractive}.

Things are different in the presence of a cosmological constant and the above 
connection between geometry and evolution gets lost. In fact, both eternal 
expansion 
and recollapse can occur for any kind of geometry, depending on the actual value 
of the vacuum energy density ${\rho}_{\Lambda}$ and of the energy 
density $\rho_M$ of ordinary stuff (matter and radiation)~\cite{kn:FI1986}. A 
striking manifestation of the cosmological constant is that the cosmic evolution 
can be {\it accelerated}. This comes about today~\footnote{In the present 
Universe, the energy density of radiation is negligible with respect to that of 
nonrelativistic matter.} for ${\rho}_{\Lambda} > 0$ and $\rho_M < 2 \, 
{\rho}_{\Lambda}$, owing to eq. (\ref{co18}). Accordingly, gravity 
becomes effectively {\it repulsive} on cosmic scales when the negative pressure of 
the vacuum dominates.

In the following, we will take for the present value of the Hubble parameter 
-- the {\it Hubble constant} -- $H_0 \simeq 70 \, km \, s^{- 1} \, Mpc^{- 1}$. 
Accordingly, the critical density is ${\rho}_c \simeq 0.92 \cdot 10^{- 29} \, g \, 
cm^{- 3} \simeq 1.36 \cdot 10^{11} \, M_{\odot} \, Mpc^{- 3}$~\footnote{The subfix 
$\odot$ denotes solar quantities. We recall that $1 \, pc \simeq 3.1 \cdot 10^{18} 
\, cm$.}. Furthermore -- for future needs --  we introduce the notations 
\begin{equation}
{\Omega}_M \equiv \frac{{\rho_M}}{{\rho}_c}~,   \label{Zco4c1}   
\end{equation}            
\begin{equation}
{\Omega}_{\Lambda} \equiv \frac{{\rho}_{\Lambda}}{\rho_c}~.   \label{Zco4c2}   
\end{equation}
It goes without saying that ${\Omega}_M$ and ${\Omega}_{\Lambda}$ represent the 
contributions to the cosmic density parameter from ordinary stuff and from the 
vacuum (respectively), and obviously  
\begin{equation}
\Omega = {\Omega}_M + {\Omega}_{\Lambda}~.   \label{Zco4c3}   
\end{equation}
Finally, all contributions to the $\Omega$ parameter considered below refer to 
the {\it present}.

The main goal of the subsequent analysis is the observational determination of 
the contributions to the $\Omega$ parameter from the various cosmic structures. 

A very useful concept for that purpose is the {\it mass-to-light} ratio $\Upsilon$ 
of a given astronomical object, having {\it total mass} $M$ and absolute {\it 
optical luminosity} $L$~\footnote{Throughout, optical luminosity generally refers 
to the blue band.}. More precisely, $\Upsilon$ is defined in solar units as
\begin{equation}
\Upsilon \equiv \frac{M/M_{\odot}}{L/L_{\odot}}~.   \label{st1}
\end{equation}
Hence, $\Upsilon$ quantifies the {\it total mass} in terms of the emitted light. 
Occasionally, we will also be concerned with a similar quantity pertaining however 
to {\it luminous mass} and denoted by ${\Upsilon}_*$, but we shall always 
carefully state which mass is being 
referred to. Of course, the Sun has $\Upsilon = 1$ by definition. Because 
main-sequence stars are characterized by the relation $L \sim M^{3.5}$, bright 
$O$, $B$ stars have $\Upsilon \ll 1$, whereas red dwarfs have $\Upsilon \gg 1$. 
Models of stellar evolution in galaxies allow for the determination of the 
mass-to-light ratio ${\Upsilon}_{*,X}$ corresponding to {\it luminous} matter in 
galaxies of various types $X$ (ellipticals $E$, lenticulars $S0$, spirals $Sa$, 
$Sb$, $Sc$ and irregulars $Irr$). The resulting mean values are displayed in Table 
1.
\vskip 0.3 cm 
Table 1
\vskip 0.2 cm
\begin{center}
\begin{tabular}{|c|c|}\hline
 &  ${\Upsilon}_{*,X} $   \\
\hline

$E$  & $6.5 $ \\
$S0$ & $5 $ \\
$Sa$ & $3 $  \\
$Sb$ & $2 $  \\
$Sc$ & $1 $  \\      
$Irr$  & $1 $  \\

\hline
\end{tabular}
\end{center}
\vskip 0.2 cm

As an important preliminary step in the search for dark matter, we estimate the 
contribution to the $\Omega$ parameter from {\it luminous} matter in galaxies. 
Actually, it turns out that the whole optical luminosity of the Universe is 
produced by galaxies, and hence such a contribution pertains to {\it all} the {\it 
luminous}~\footnote{We warn the reader that -- unless otherwise stated -- luminous 
refers to {\it optical} luminosity.} matter in the Universe. Let us begin by 
recalling that galaxy surveys allow for the determination of the {\it luminosity 
function} ${\Phi}(L)$, which gives the mean number density of galaxies per unit 
luminosity. Its analytic expression is
\begin{equation}
{\Phi}(L)= \frac{{\Phi}_*}{L_*}  \left( \frac{L}{L_*} \right)^{\alpha} \, 
e^{- L/L_*}~,   \label{1uuu1}     
\end{equation}
where ${\Phi}_* \simeq 0.41 \cdot 10^{- 3} \, Mpc^{- 3}$, $L_* \simeq 2.53 \cdot 
10^{10} \, L_{\odot}$ and $\alpha \simeq - \, 1.25$~\cite{kn:LONGAIR}. Because the 
whole optical 
luminosity of the Universe arises from galaxies, the mean cosmic luminosity 
density is
\begin{equation}
{\cal L} = \int_0^{\infty} dL ~ L \, {\Phi}(L) \simeq 1.4 \cdot 10^8 \, 
L_{\odot} \, Mpc^{- 3}~.   \label{2uuu1}     
\end{equation}
Observations also provide the fraction ${\cal F}_X$ of the cosmic luminosity 
produced by the galaxy population of type $X$~\cite{kn:FHP1998}. Their values 
are reported in Table 2.
\vskip 0.3 cm 
Table 2
\vskip 0.2 cm
\begin{center}
\begin{tabular}{|c|c|}\hline
 &  ${\cal F}_X $   \\
\hline

$E$  & $0.11 $ \\
$S0$ & $0.21$ \\
$Sa$ & $0.28 $  \\
$Sb$ & $0.29 $  \\
$Sc$ & $0.05 $  \\      
$Irr$  & $0.06 $  \\

\hline
\end{tabular}
\end{center}
\vskip 0.2 cm
Hence, the corresponding mean luminosity density is
\begin{equation}
{\cal L}_X = {\cal F}_X  \, {\cal L}  \simeq 1.4 \cdot 10^8 \, {\cal F}_X \, 
L_{\odot} \, Mpc^{- 3}~.   \label{3uuu1}     
\end{equation}
We can convert ${\cal L}_X$ into the mean cosmic density ${\rho}_{*,X}$ of 
{\it luminous} mass contained in the galaxy population $X$ by the mass-to-light 
ratio ${\Upsilon}_{*,X}$, thereby getting
\begin{equation}
{\rho}_{*,X} = {\cal L}_X \, {\Upsilon}_{*,X} M_{\odot} \, L_{\odot}^{- 1} \simeq 
1.4 \cdot 10^8  \, {\cal F}_X \, {\Upsilon}_{*,X} \, M_{\odot} \, Mpc^{- 3}~.   
\label{5uuu1}     
\end{equation}
Consequently, the resulting contribution to $\Omega$ is
\begin{equation}
{\Omega}_{*,X} \equiv \frac{{\rho}_{*,X}}{{\rho}_c}  \simeq 1.03 
\cdot 10^{- 3} \, 
{\cal F}_X \, {\Upsilon}_{*,X}~,    \label{A6uuu1}     
\end{equation}
whose values -- for the various galaxy populations -- are listed in Table 3.
\vskip 0.3 cm 
Table 3
\vskip 0.2 cm
\begin{center}
\begin{tabular}{|c|c|}\hline
 &  $ {\Omega}_{*,X} $   \\
\hline

$E$  & $0.73 \cdot 10^{- 3} $ \\
$S0$ & $1.08 \cdot 10^{- 3} $ \\
$Sa$ & $0.87 \cdot 10^{- 3} $  \\
$Sb$ & $0.59 \cdot 10^{- 3} $  \\
$Sc$ & $0.05 \cdot 10^{- 3} $  \\      
$Irr$  & $0.06 \cdot 10^{- 3} $  \\

\hline
\end{tabular}
\end{center}
\vskip 0.2 cm
Thus, we come to the conclusion that the contribution from {\it all} the {\it 
luminous} mass in the Universe is 
\begin{equation}
{\Omega}_* = \sum_X {\Omega}_{*, X} \simeq 0.004~.   \label{mo3}   
\end{equation}

As a matter of fact, the above argument -- relating the mass-to-light ratio of a 
galaxy population ${\Upsilon}_{*,X}$ to the corresponding contribution to the 
cosmic density parameter ${\Omega}_{*,X}$ -- ramains true even if the 
mass-to-light ratio ${\Upsilon}_X$ is considered, which refers to {\it all} 
galactic mass and not just to the luminous component. Accordingly, the counterpart 
of eq. (\ref{A6uuu1}) yields the contribution ${\Omega}_X$ from the galaxy 
population $X$. Explicitly
\begin{equation}
{\Omega}_X \simeq 1.03 \cdot 10^{- 3} \, {\cal F}_X \, {\Upsilon}_X~.    
\label{KA6uuu1}     
\end{equation}
Of course, at this stage $\Upsilon_X$ is {\it unknown}, owing to the existence of 
dark matter in galaxies. Nevertheless -- thanks to eq. (\ref{KA6uuu1}) -- the 
determination of $\Upsilon_X$ (to be discussed in Sect. 3) will allow us to 
quantify the {\it cosmological relevance} of a given galaxy population.

\section{Astrophysical evidence of dark matter}

This Section is the core of the present review and addresses the observational 
evidence for dark matter in galaxies and in clusters of galaxies.

\subsection{Spiral galaxies}

Stars in spiral galaxies are mainly contained -- along with cold neutral hydrogen 
$HI$ clouds -- in a thin {\it disk} characterized by an exponential surface 
brightness profile
\begin{equation}
I(R) \sim e^{- R/R_d}~,   \label{gn5}   
\end{equation}
where $R$ is the galactocentric distance and $R_d$ denotes the {\it disk scale 
lenght}. Typically, one finds $R_d \simeq 2 - 4 \, kpc$ and the disk {\it optical 
radius} turns out to be $R_{opt} \simeq 4 \, R_d$. Stars and cold $HI$ clouds 
travel on nearly {\it circular} orbits around the galactic centre with velocity 
$v_c(R)$, and so their centripetal acceleration equals the gravitational one
\begin{equation}
\frac{v_c^2(R)}{R} = - \, g_R(R,0)~. \label{s4}
\end{equation}
Hence, the {\it rotation curve} -- namely the graph of $v_c(R)$ versus $R$ -- 
traces the gravitational acceleration in the disk $g_R(R,0)$. This fact lies at 
the basis of the best strategy to discover dark matter in spiral galaxies. 

Basically, a rotation curve is constructed by measuring the circular velocity -- 
at different values of the radius -- by the Doppler shift of certain spectral lines 
in either the optical or the radio band of the emitted galactic radiation. The {\it 
observed} rotation curve of a given spiral is then compared with the one produced 
by {\it luminous} matter alone: a discrepancy would be a clear signal of dark 
matter. 

Let us consider the main steps of this procedure in some detail. In spite of the 
fact that it is virtually impossible to find identical rotation curves, it turns 
out that they all share the {\it same} qualitative behaviour: observations show 
that they {\it rise linearly} in the inner region until a maximum is reached near 
$R \simeq 2 \, R_d$, beyond which they stay {\it flat} out to the last measured 
point. Schematically
\begin{equation}
v_c(R) \sim \left\{ \begin{array}{ll}
R~, & R < R_d \\
constant~, & R > 3 \, R_d~. 
\end{array}
\right. \label{Zaa28}
\end{equation}
As we said, such a rotation curve has to be compared with the one arising solely 
from {\it luminous} matter, whose evaluation proceeds as follows. In the first 
place, the surface brightness profile -- as given by eq. (\ref{gn5}) with $R_d$ 
fixed by a photometric fit -- has to be converted into the disk surface density 
profile ${\Sigma}(R)$. Because colour and luminosity gradients in spiral disks are 
generally modest, it makes sense to suppose that the disk mass-to-light ratio 
${\Upsilon}_d$ is {\it constant}. Accordingly, one gets ${\Sigma}(R) \sim 
{\Upsilon}_d \, I(R)$, and so eq. (\ref{gn5}) entails
\begin{equation}
{\Sigma}(R) \sim e^{- R/R_d}~,   \label{Mgn5}   
\end{equation}
It can next be shown that this mass distribution produces the following rotation 
curve 
\begin{equation}
v_c (R) \sim \left[ I_0\left(\frac{R}{2R_d} \right) K_0\left(\frac{R}{2R_d} 
\right) - I_1\left(\frac{R}{2R_d} \right) K_1\left(\frac{R}{2R_d} \right) \right]
^{1/2} R~,  \label{s11}
\end{equation}
where $I_0(\cdot)$, $I_1(\cdot)$, $K_0(\cdot)$ and $K_1(\cdot)$ are modified 
Bessel functions~\cite{kn:BT1987}. Although eq. (\ref{s11}) looks complicated, its 
qualitative behaviour is very simple: a {\it linear rise} in the inner region 
continues until a maximum is reached near $R \simeq 2 \, R_d$, which is followed 
by a {\it keplerian fall-off} at larger galactocentric distances. To a good 
approximation, eq. (\ref{s11}) can be rewritten as 
\begin{equation}
v_c(R) \sim \left\{ \begin{array}{ll}
R~, & R < R_d \\
R^{- 1/2}~, & R > 3 \, R_d~. 
\end{array}
\right. \label{aa28}
\end{equation}
Let us now compare eqs. (\ref{Zaa28}) and (\ref{aa28}). As far as the inner 
region $R < R_d$ is concerned, the agreement is good, thereby implying that 
luminous matter is the whole story. But in the outer region $R > 3 \, R_d$ the 
disagreement is dramatic, with the circular velocity systematically 
larger than expected on the basis of luminous matter alone. Actually, a 
larger $v_c$ (for fixed $R$) implies by eq. (\ref{s4}) a larger $g_R$, 
which entails in turn by Poisson equation (see below) a larger mass density 
${\rho}$. Consequently, {\it dark} matter {\it must} lurk at galactocentric 
distances $R > 3 \, R_d$.

Clearly, the flat behaviour of the observed rotation curves provides 
solid evidence that the outer region of spiral galaxies is dominated by dark 
matter. This turns out to be a {\it universal} properties of spiral galaxies. 

Yet, the actual shape of the dark matter distribution {\it cannot} be unambiguously 
determined from the rotation curve alone. This is true even under the simplifying 
assumption 
that such a distribution -- just like the one of luminous matter -- is {\it 
axisymmetric} about the galaxy rotation axis. Employing cylindrical coordinates 
$(R, \phi, z)$, the gravitational acceleration in a generic point ${\bf g}(R, z)$ 
is related to the mass density $\rho(R, z)$ by {\it Poisson equation}
\begin{equation}
\frac{1}{R} \frac{\partial }{\partial R} \Bigl( R \, g_R(R, z) \Bigr) + 
\frac{\partial g_z (R, z)}{\partial z}  = - ~ 4 \pi G \, \rho(R, z)~. \label{s5}
\end{equation}
Owing to eq. (\ref{s4}), the rotation curve merely fixes $g_R(R, 0)$, and so the 
lack of knowledge about $g_R(R, z)$ with $z \neq 0$ and $g_z(R, z)$ prevents the 
unique determination of $\rho(R, z)$. {\it Only} by assuming that the dark matter 
distribution has {\it spherical symmetry} (about the galaxy centre) can the dark 
matter density profile $\rho(r)$ be {\it uniquely} derived from the rotation curve. 
Indeed, now the Poisson equation reads~\footnote{Notationally, we denote by $r$ or 
$\cal R$ the radius of a sphere, while $R$ stands for the radius of a circle.}
\begin{equation}
\frac{1}{r^2} \frac{d}{dr} \left( r^2 g_r (r) \right) = - \, 4 \pi G \rho~.   
\label{Ks5}
\end{equation}
So, upon integration we get
\begin{equation}
g_r (r) = - \, \frac{G \, M( < r)}{r^2}~,   \label{Js5}
\end{equation}
where
\begin{equation}
M(< r) \equiv 4 \pi \int_0^r dr' r'^2 \rho(r')   \label{g7}
\end{equation}
denotes the {\it integrated mass profile}, namely the total mass inside the sphere 
of radius $r$. Combining eqs. (\ref{s4}) and (\ref{Js5}) together, we find
\begin{equation}
M( < r) \sim v^2_c (r) \, r~.   \label{Zg7}
\end{equation}
Still, in the region where the observed rotation curve is {\it flat} eq. 
(\ref{Zg7}) becomes
\begin{equation}
M( < r) \sim  r~,   \label{ZZg7}
\end{equation}
and so eq. (\ref{g7}) implies
\begin{equation}
{\rho} (r) \sim r^{- 2}~,   \label{KZZg7}
\end{equation}
which is the density profile of a {\it singular isothermal sphere} 
(SIS)~\footnote{A SIS is a self-gravitating spherical model with {\it diagonal} 
pressure tensor and velocity dispersion {\it independent} of 
position~\cite{kn:BT1987}.}. Thus, 
we come to the conclusion that spiral galaxies are 
surrounded by a SIS halo {\it dominated} by {\it dark} matter~\footnote{However, 
one should not forget that such a conclusion rests upon the simple but unproved 
assumption of spherical symmetry.}.

The systematic analysis of spiral galaxy rotation curves started nearly twenty 
years ago, both in the optical and in the radio band, where the $21 \, cm$ 
emission line of $HI$ is used~\footnote{This line arises from the hyperfine 
transitions in the ground state. Notice that the higher level is populated because 
the temperature of the interstellar medium is definitely larger than the energy 
gap between the hyperfine levels.}. 

Early optical studies~\cite{kn:RUBIN} found that the mass-to-light ratio 
${\Upsilon}_{opt}$ 
pertaining to a sphere with radius $R_{opt}$ systematically exceeds by 
roughly a factor of $2$ the mass-to-light ratio ${\Upsilon}_*$ of the luminous mass 
$M_*$. Because ${\Upsilon}_{opt}/{\Upsilon}_* = M( < R_{opt})/M_*$, it follows
\begin{equation}
M_{dark} ( < R_{opt}) \simeq M_*~,   \label{KJZZg7}
\end{equation}
meaning that the optical region of a spiral contains roughly {\it equal} amounts 
of {\it luminous} and {\it dark} matter. 

However, such a conclusion should be understood more like a suggestion than a 
real proof. For, the optical method is (just by definition) bound to probe the 
region $R < 4 \, R_d$ (recall that $R_{opt} \simeq 4 \, R_d$), and in fact the 
above analysis has been carried out up to $R \simeq 3.5 \, R_d$. Still, eq. 
(\ref{s11}) implies that at $R \simeq 3.5 \, R_d$ the circular velocity of 
luminous matter has decreased only by $8 \, {\%}$ relative to its maximum at 
$R \simeq 2 \, R_d$. So, it is very difficult to rule out the keplerian fall-off 
by restricting the attention to such a narrow range of galactocentric distances.

Remarkably enough, radio observations resolve the issue. Indeed, $HI$ clouds 
typically extend out to {\it twice} the optical radius, thereby allowing for the 
determination of $v_c (R)$ up to $R \simeq 8 \, R_d$. In this way, the existence 
of dark matter can be established even outside the optical region, where it 
actually dominates the mass distribution. A beautiful example concerns the $Sc$ 
spiral galaxy $NGC 3198$, whose rotation curve has been mapped out to $R \simeq 10 
\, R_d \simeq 30 \, kpc$~\cite{kn:vbbs1985}. The resulting mass-to-light ratio is 
${\Upsilon} \simeq 
18$. We know from Table 1 that for an $Sc$ spiral the mass-to-light ratio of 
luminous matter is ${\Upsilon}_{*,Sc} \simeq 1$, and so the following result emerges
\begin{equation}
M_{dark} ( < 30 \, kpc) \simeq 17 \, M_*~,   \label{ZJZZg7}
\end{equation}
with $M_*$ denoting the luminous mass.

At still larger galactocentric distances, no stars or cold $HI$ clouds are present. 
Therefore, tracers of a different kind have to be identified in order to probe the 
mass profile of dark halos. Several bright spirals happen to possess fainter -- and 
presumably less massive -- satellite galaxies (similarly to the case of our Milky 
Way, which has the two Magellanic Clouds as satellite galaxies). This fact offers 
the possibility to investigate the gravitational field of the primary galaxy 
through the dynamical behaviour of the satellites. Of course, their orbital period 
is by far too long to observe a significant portion of the orbit, and so the 
ensemble of satellites has to be handled in a statistical fashion~\cite{kn:BT1981}. 
This strategy 
has been applied to a sample of 115 satellites around 69 primaries, having mean 
luminosity $L \simeq 2 \cdot 10^{10} \, L_{\odot}$~\cite{KN:Zw1994}. We 
stress that the underlying philosophy is to suppose that the primaries are 
sufficiently similar that the satellites can be treated as orbiting a {\it single} 
(typical) galaxy, thereby significanty enhancing the statistical relevance of the 
satellite sample. This analysis entails that dark halos extend beyond $R \simeq 
200 \, kpc$, with
\begin{equation}
M ( < 200 \, kpc) \simeq 2 \cdot 10^{12} \, M_{\odot}~.   \label{AJZZg7}
\end{equation}
Then the corresponding mass-to-light ratio is ${\Upsilon} \simeq 100$. Because 
this relation can be regarded as {\it typical} for spiral galaxies, we can state 
that the mean mass-to-light ratio of these galaxies is
\begin{equation}
{\Upsilon}_S \simeq 100~.   \label{ZZs35}    
\end{equation}
Recalling the values quoted in Table 1, it follows that {\it any} spiral obeys the 
condition
\begin{equation}
M_{dark} > 30 \, M_*~,   \label{ZZZs35}    
\end{equation}
thereby implying that all spiral galaxies are totally {\it dominated} by {\it 
dark} matter.

A comparison among observations at different galactocentric distances of the same 
spiral and for different spirals yields~\cite{kn:BLD1995}
\begin{equation}
{\Upsilon}_S (r) \simeq 60 \left( \frac{r}{100 \, kpc} \right)~.   \label{s36}    
\end{equation}

Let us finally address the {\it cosmological relevance} of spiral galaxies. We 
consider first the contribution to the $\Omega$ parameter from {\it luminous} 
matter alone. Recalling the relevant values listed in Table 3, we get
\begin{equation}
{\Omega}_{*,S} \simeq 1.5 \cdot 10^{- 3}~.   \label{s36AA}    
\end{equation}
Because of the presence of dark matter, the {\it total} contribution to $\Omega$ 
turns out to be much larger. By combining eq. (\ref{KA6uuu1}) and (\ref{ZZs35}) 
together and 
using the relevant values reported in Table 2, we find
\begin{equation}
{\Omega}_S \simeq  1.03 \cdot 10^{- 1} \Bigl( {\cal F}_{Sa} + {\cal F}_{Sb} + 
{\cal F}_{Sc} \Bigr) \simeq 6.4 \cdot 10^{- 2}~.    \label{s36AB}     
\end{equation}

\subsection{Elliptical galaxies}

Luminous matter in elliptical galaxies has a {\it spheroidal} distribution, well 
described by the {\it De Vaucouleurs} surface brightness profile
\begin{equation}
I(R) \sim  \exp \left\{-7.67 \left[ (R/R_e)^{1/4} - 1 \right] \right\}~,
\label{p2ab}
\end{equation}
where $R$ is the galactocentric distance and $R_e$ is the {\it effective 
radius} (typically $R_e \simeq 3 - 5 \, kpc$). Moreover, the star motion in 
ellipticals is highly chaotic, with velocity dispersions usually as large as 
velocities themselves. Manifestly, in such a situation a rotation curve provides 
{\it no} information about the galactic gravitational field, and so different 
techniques have to be devised to look for dark matter in elliptical galaxies.

A classic approach rests upon the {\it dynamical analysis} of stellar motion and 
can be summarized as follows. Any {\it specific} stellar population of ellipticals 
can be thought of as a collisionless fluid in a steady state, resulting from the 
balance between the kinetic pressure -- brought about by the above-mentioned 
chaotic motion -- and the overall gravitational field. Assuming spherical 
symmetry and denoting by $r$ the galactocentric distance, it can be shown that the 
star number density profile $n_s(r)$ obeys the following 
equation~\cite{kn:BT1987}
\begin{equation}
\frac{d}{dr} \Bigl( n_{s} \sigma^2_{r} \Bigr) + \frac{2 \, n_{s} A \, 
\sigma^2_{r}}{r} + \frac{G M(< r) \, n_{s}}{r^2} = 0~.   \label{ee5}
\end{equation} 
This is just {\it Euler equation} for a fluid with {\it nondiagonal} pressure tensor 
parametrized by the radial velocity dispersion ${\sigma}_r$ and the anisotropy 
function $A(r)$. We emphasize that $M( < r)$ -- as defined by eq. (\ref{g7}) -- 
pertains to the {\it total} galactic mass (responsible for the overall 
gravitational field). It is very easy to see that eq. (\ref{ee5}) can be rewritten 
in the form
\begin{equation}
M(< r) = - \frac{\sigma^2_{r}}{G} \left( \frac{d \ln n_{s}}{d \ln r} + 
\frac{d \ln \sigma^2_{r}}{d \ln r} + 2\, A \right)r~.   \label{ee6}
\end{equation}
So, we can find the overall integrated mass profile of the elliptical in question 
provided that we succeed in determining the three functions $n_s(r)$, 
${\sigma}_r(r)$ and $A(r)$. From a conceptual point of view, $M( < r)$ plays here 
the same r\^ole 
as $v_c(R)$ did for spiral galaxies: a discrepancy between $M( < r)$ and 
the integrated mass profile of {\it luminous} matter alone would provide positive 
evidence for dark matter. Unfortunately, only 
the surface brightness profile on the sky $I(R)$ and the velocity dispersion 
profile along the line-of sight $\sigma_{\parallel}(R)$ are the available {\it 
observables}, and so there is not enough information to uniquely fix the unknown 
functions $n_s(r)$, ${\sigma}_r(r)$ and $A(r)$. As a result, $M( < r)$ can be 
determined only by making some assumption on the functional form of $A(r)$, as 
suggested by models of galaxy formation~\cite{kn:SAGLIA}. For a long time, 
instrumental limitations prevented the application of such a dynamical analysis to 
a tracer population having a sufficiently large galactocentric distance, thereby 
severely hindering its effectiveness. But in the last few years the situation has 
considerably improved and today globular clusters and planetary nebulae can be 
mapped out to $r \simeq 6 \, R_e$. As far as dark matter is concerned, the above 
dynamical analysis leads to a result which {\it strongly depends} on the specific 
elliptical that is being considered. In some cases, there is {\it no} evidence for 
dark matter out to $r \simeq 4 \, R_e$, whereas is other cases one typically gets 
${\Upsilon}_E \simeq 10 - 15$~\cite{kn:GERHARD}. Recalling from Table 1 that 
${\Upsilon}_{*,E} \simeq 6.5$, it follows
\begin{equation}
M_{dark} ( < 4 \, R_e) \simeq M_*~,   \label{NTee6}
\end{equation}
where again $M_*$ denotes the luminous mass. Quite recently, this conclusion has 
been confirmed by a totally different technique, namely by the observation of {\it 
strong gravitational lensing}~\footnote{This phenomenon will be discussed later 
on.} of a background galaxy produced by the elliptical $MG2016 + 112$~\cite{TREU}. 
Thus -- analogously to what happens for spiral galaxies -- also the optical region 
of ellipticals contains roughly {\it equal} amounts of {\it luminous} and 
{\it dark} matter. However, there are several {\it exceptions} to this statement 
(which is {\it not} the case for spiral galaxies).

Bright elliptical galaxies generally contain a sizable amount of hot ionized 
gas at temperature $T_g \simeq 3 \cdot 10^6 - 1 \cdot 10^7 \, K$, which gives rise 
-- by thermal Bremsstrahlung (free-free transitions) -- to an $X$-ray emission 
with luminosity $L_X \simeq 10^{39} - 10^{42} \, erg \, s^{-1}$. Such an $X$-ray 
emission is considerably more diffuse than optical light, implying that the gas 
distribution extends out to $r \simeq 20 - 80 \, kpc$~\cite{kn:FJT1985}. Quite 
remarkably, the previous method can be applied -- without the above-mentioned 
shortcomings -- to this gas, whose pressure tensor is necessarily {\it diagonal} 
(because of the {\it Pascal law}). Consequently, the anisotropy function $A(r)$ 
vanishes identically and eq. (\ref{ee6}) becomes 
\begin{equation}
M(< r) = - \frac{k_B T_g}{G m} \left( \frac{d \ln n_g}{d \ln r} + 
\frac{d \ln T_g}{d \ln r} \right)r~, \label{ee13bis}   
\end{equation}
where $n_g(r)$ and $T_g(r)$ are the gas density and temperature profiles 
(respectively), while $m \simeq 0.6 \, m_p$ denotes the mean particle mass of the 
gas. Of course, eq. (\ref{ee13bis}) rests upon the assumption -- which will be 
justified later on -- that the gas is in {\it hydrostatic equilibrium}.

The present strategy can be implemented as follows. $X$-ray observations -- 
performed by satellite-borne detectors -- provide the surface brightness profile 
of a given bright elliptical. An excellent fit to the data is achieved by the 
analytic expression
\begin{equation}
I_X(R) \simeq \left[ 1 + \left( \frac{R}{a_X} \right)^2 \right]^{-3 \beta + 
1/2}~,   \label{ee14}   
\end{equation}
where typically $\beta \simeq 0.4 - 1.0$ and $a_X \simeq 1 - 9 \, kpc$. Upon 
deprojection, eq. (\ref{ee14}) yields the $X$-ray luminosity density
\begin{equation}
j_X(r) \sim \left[ 1 + \left( \frac{r}{a_X} \right)^2 
\right]^{- 3 \beta}~.
\label{ee15a}
\end{equation}
Determining the gas temperature profile is a much more difficult job, but to a 
good approximation it can be assumed that the gas distribution is {\it 
isothermal}. Because the $X$-ray emission is due to Bremsstrahlung, the luminosity 
density $j_X(r)$ goes like the {\it square} of the gas number density $n_g(r)$, 
and so from eq. (\ref{ee15a}) we get
\begin{equation}
n_g(r) \sim \left[ 1 + \left( \frac{r}{a_X} \right)^2 \right]^{- 3 
\beta/2}~,   \label{ee17}    
\end{equation}
which -- upon substitution into eq. (\ref{ee13bis}) -- gives
\begin{equation}
M(< r) = \frac{3 \beta k_B T_g}{G m} \frac{(r/a_X)^2}{1 + (r/a_X)^2} \, r~.   
\label{ee18}
\end{equation}

What is the physical meaning of this result? We pointed out that typically one 
finds $a_X < 10 \, kpc$. Thus, in the region $r > 10 \, kpc$ -- which corresponds 
to the galactic {\it halo} -- eq. (\ref{ee18}) takes the form
\begin{equation}
M(< r) \sim r~,   \label{ee19}  
\end{equation}
from which -- thanks to eq. (\ref{g7}) -- the corresponding overall density 
profile is immediately deduced and reads
\begin{equation}
\rho(r) \sim r^{ - 2}~.  \label{ee19aa}  
\end{equation}
We already encountered this expression and we know that it describes a {\it 
singular isothermal sphere} (SIS). So, we see that bright ellipticals possess a 
SIS {\it halo}~\footnote{This conclusion rests on the hydrostatic equilibrium 
assumption, which can be justified as follows. Basically, the dynamical behaviour 
of the hot gas depends on its ability to get rid of its thermal energy and is 
determined by the competition between the {\it cooling time} $t_{cool}$ and the 
{\it free-fall time} $t_{ff}$. When $t_{cool} < t_{ff}$, cooling occurs efficiently 
and the gas collapses toward the galactic centre. On the other hand, for $t_{cool} 
> t_{ff}$ the gas stays in hydrostatic equilibrium. It can be shown that -- for the 
observed values of the relevant parameters (see below) -- just the latter 
situation is actually realized outside the central region of a bright elliptical.}.

What is such an halo made of? In order to settle this issue, the integrated mass 
profile of the gas $M_g(<r)$ has to be computed and compared with $M( < r)$ as 
given by eq. (\ref{ee18}), for the observed {\it mean} values of the relevant 
parameters $T_g \simeq 7.8 \cdot 10^6 \, K$, $\beta \simeq 0.5$ and $a_X \simeq 5 
\, kpc$~\cite{kn:OPC2003}. Once again, a discrepancy between $M_g(<r)$ and 
$M( < r)$ would signal the presence of dark matter. An explicit calculation 
yields
\begin{equation}
M_g(< r) \simeq 6.3 \cdot 10^7 \left( \frac{r}{kpc} \right)^{1.5} 
M_{\odot}~,  \label{ee20aa}   
\end{equation}
and
\begin{equation}
M(< r) \simeq 3.6 \cdot 10^{10} \left( \frac{r}{kpc} \right) M_{\odot}~.   
\label{ee20}
\end{equation}
Thus, we conclude that bright elliptical galaxies are surrounded by a SIS halo 
{\it dominated} by {\it dark} matter. We stress that -- in spite of the great 
difference in their optical properties -- bright ellipticals and spirals are 
qualitatively identical as far as dark matter is concerned. Moreover, assuming 
$L \simeq 2 \cdot 10^{10} \, L_{\odot}$ for the mean optical luminosity of bright 
ellipticals, their mean mass-to-light ratio resulting from eq. (\ref{ee20}) is  
\begin{equation}
{\Upsilon}_E (r) \simeq 180 \left( \frac{r}{100 \, kpc} 
\right)~.   \label{ZZs36}    
\end{equation}
Comparing eqs. (\ref{s36}) and (\ref{ZZs36}), we see that -- for equal values of 
radius and luminosity -- bright ellipticals contain roughly 3 times more dark 
matter than spirals.

How big are the halos of elliptical galaxies? Preliminary observations based on 
statistical gravitational lensing~\cite{kn:SLE} show -- though with large 
uncertainties -- that bright ellipticals have halos similar to those of spirals, 
which seem to be consistent with eq. (\ref{ZZs36}). Hence, the resulting mean 
value of the mass-to-light ratio for these galaxies is
\begin{equation}
{\Upsilon}_{E,b} \simeq 300~.   \label{ZZs36AA}    
\end{equation}
Combining this result with ${\Upsilon}_{*,E} \simeq 6.5$ (from Table 1), we get
\begin{equation}
M_{dark} \simeq 45 \, M_*   \label{ZZs36AB}    
\end{equation}
which implies that bright elliptical galaxies are {\it totally} dominated by {\it 
dark} matter. As already emphasized, a similar clear-cut statement {\it cannot} be 
made for fainter ellipticals.

Let us finally consider the {\it cosmological relevance} of elliptical galaxies. 
As far as their {\it luminous} matter is concerned, we recall from Table 3 that 
the contribution to the $\Omega$ parameter is
\begin{equation}
{\Omega}_{*,E} \simeq 0.73 \cdot 10^{- 3}~.   \label{ntZZs36}    
\end{equation}
Obviously, the presence of dark matter makes the {\it total} contribution to the 
$\Omega$ parameter considerably larger, but the lack of knowledge about the 
relevance of dark matter in fainter ellipticals prevents a reliable estimate. The 
best we can do is to derive the {\it upper bound}
\begin{equation}
{\Omega}_E < 6.5 \cdot 10^{- 2}~,   \label{ntZZs36CC}    
\end{equation}
which rests upon eqs. (\ref{KA6uuu1}) and (\ref{ZZs36AA}).

\subsection{Clusters of galaxies}

Galaxies are not randomly distributed throughout the Universe, but tend to 
aggregate on the Megaparsec scale. Rich clusters contain $30 - 300$ galaxies 
inside a sphere of {\it Abell radius}, conventionally defined as ${\cal R}_A 
\simeq 2.1 \, Mpc$. The resulting optical luminosity is $L ( < {\cal R}_A) 
\simeq 1.2 \cdot 10^{13} - 1.2 \cdot 10^{14} \, L_{\odot}$. {\it Regular} clusters 
typically show a centrally condensed region and are characterized by spherical 
symmetry, whereas {\it irregular} clusters have no characteristic shape.

As already pointed out, the first evidence for a large amount of dark matter in 
the Universe came from the virial analysis of the Coma cluster by Zwicky in 1933. 
This strategy has since been applied to many regular clusters and can be 
summarized as follows. Any isolated self-gravitating system reaches an equilibrium 
state, in which gravity is balanced by kinetic pressure. Accordingly, the 
potential energy $U$ is related to the kinetic energy $K$ by the {\it virial 
theorem}
\begin{equation}
2 K + U = 0~.   \label{nt1}   
\end{equation}
Assuming spherical symmetry, the potential energy can be represented as
\begin{equation}
U = -  \frac{\alpha G M^2}{\cal R}~,   \label{g10}
\end{equation}
where $M$ is the total mass, $\cal R$ is the radius of the system and $\alpha$ is 
a constant which reflects the actual density profile. Because the kinetic energy 
can be written -- in terms of the mean-square velocity $\langle v^2 \rangle$ -- as 
\begin{equation}
K = \frac{1}{2} M \langle v^2 \rangle   \label{nt2}   
\end{equation}
eq. (\ref{nt1}) yields
\begin{equation}
M = \frac{ \langle v^2 \rangle \cal R}{\alpha G}~.   \label{nt3}   
\end{equation}
However, the application of eq. (\ref{nt3}) to regular clusters is not as 
straightforward as it might seem. For, the cluster density profile is unknown, and 
so the constant $\alpha$ in eq. (\ref{nt3}) gives rise to an uncertainty in the 
mass determination. Another source of uncertainty is due to the fact that real 
clusters do not possess a sharp edge, making $\cal R$ an ill-defined quantity. A 
way out of both difficulties is provided by the following alternative version of 
eq. (\ref{nt3})
\begin{equation}
M = \frac{ \langle v^2 \rangle R_M}{G}~,   \label{nt4}   
\end{equation} 
where $R_M$ is the {\it mean effective radius}, a quantity which is 
observationally well-defined in terms of galaxy counts~\cite{kn:RPK1972}. Notice 
that $\alpha$ has disappeared from eq. (\ref{nt4}). Still, an additional 
uncertainty in the estimated mass arises through the determination of $\langle v^2 
\rangle$ for the cluster galaxies. For, what is really measured is the 
one-dimensional galaxy velocity dispersion $\sigma$ along 
the line of sight, but there is no way to tell how $\sigma$ is related to $\langle 
v^2 \rangle$ in the general case. So, it is usually assumed that the galaxy 
velocity distribution in regular clusters is {\it isotropic}, in which case 
$\langle v^2 \rangle = 3 {\sigma}^2$. In spite of this and other uncertainties, 
it is widely believed that the virial mass estimates for regular clusters are 
fairly good. A sample of values of the mass-to-light ratio ${\Upsilon}_{RC}$ for regular 
clusters derived by the virial theorem~\cite{kn:CARLBERG1995} is reported in Table 
4.

\vskip 0.3 cm 
Table 4
\vskip 0.2 cm
\begin{center}
\begin{tabular}{|c|c|c|c|}\hline
Cluster &  ${\Upsilon}_{RC} $   \\
\hline

$A2390$  & $173 $ \\
$MS0016+16$ & $202 $ \\
$MS0302+16$ & $157 $  \\
$MS0440+02$ & $218 $  \\
$MS0451+02$ & $250 $  \\
$MS0451-3$  & $275 $  \\
$MS0839+29$ & $200 $  \\
$MS0906+11$ & $560 $  \\
$MS1006+12$ & $204 $  \\
$MS1008-12$ & $154 $  \\
$MS1224+20$ & $148 $  \\
$MS1231+15$ & $123 $  \\
$MS1358+62$ & $138 $  \\
$MS1455+22$ & $412 $  \\
$MS1512+36$ & $164 $  \\
$MS1621+26$ & $106 $  \\

\hline
\end{tabular}
\end{center}
\vskip 0.2 cm

Clusters of galaxies contain a large amount of hot ionized gas at temperature 
$T_g \simeq 1 \cdot 10^7 - 1.5 \cdot 10^8 \, K$, which produces an $X$-ray emission 
with luminosity $L_X \simeq 6 \cdot 10^{42} - 2 \cdot 10^{45} \, erg \, s^{-1}$ by 
thermal Bremsstrahlung (free-free transitions)~\cite{kn:SARAZIN}. Moreover, the 
$X$-ray emission is 
more diffuse than optical luminosity, entailing that the gas distribution extends 
well beyond the Abell radius. Manifestly, the situation is largely analogous to 
what happens in bright elliptical galaxies, and so an analysis quite similar to 
the one considered in the previous Subsection can be carried out for {\it regular} 
clusters as well. Even in this case the {\it hydrostatic equilibrium} assumption 
outside the central region turns out to be justified. Furthermore, {\it 
isothermality} is a good first approximation. Therefore, by just repeating the 
same steps as before the resulting cluster integrated mass profile is
\begin{equation}
M(< r) = \frac{3 \beta k_B T_g}{G m} \frac{(r/a_X)^2}{1 + (r/a_X)^2} \, r~,   
\label{NTee18}
\end{equation}
where the observed {\it mean} values of the relevant parameters are now $T_g 
\simeq 5 \cdot 10^7 \, K$, $\beta \simeq 0.7$ and $a_X \simeq 0.3 \, Mpc$. Because 
one typically finds $a_X < 0.5 \, Mpc$, in the region $r > 0.5 \, Mpc$ eq. 
(\ref{NTee18}) becomes
\begin{equation}
M(< r) \sim r~.   \label{NTee19}  
\end{equation}
We know that the corresponding overall density profile is  
\begin{equation}
\rho(r) \sim r^{ - 2}~,  \label{NTee19aa}  
\end{equation}
which describes a {\it singular isothermal sphere} (SIS). Quite remarkably, 
observations show that not only spiral and bright elliptical galaxies but also 
regular clusters of galaxies are surrounded by a SIS {\it halo}.

While the gas is obviously a constituent of the SIS cluster halo, a nontrivial 
contribution from dark matter is present. Indeed, we know that dark matter 
dominates galaxies, and so it necessarily lurks inside regular clusters. Moreover, 
some further dark matter can exist in the intracluster space. So, the real 
question is whether hot gas or dark matter dominates the cluster mass budget. It 
goes without saying that this issue can be resolved by evaluating the gas mass 
fraction
\begin{equation}
f_g(r) \equiv \frac{M_g(< r)}{M(< r)}   \label{NTee19aa1}  
\end{equation}
for the above mean values of the parameters involved. It turns out that $f_g$ is 
{\it independent} of $r$ (this is simply due to $\beta \simeq 0.7$) and we get
\begin{equation}
f_g \simeq 0.12~.   \label{NTee19aa2}  
\end{equation}
In addition, the total gas mass comes out invariably {\it larger} than the 
luminous mass of the cluster galaxies. Thus, we conclude that also regular 
clusters of galaxies are {\it dominated} by {\it dark} matter.

Besides leading to values of the mass-to-light ratio for regular clusters which 
are {\it consistent} with those previously found from the virial analysis, this 
method also has a different implication. We will discuss in the next Section the 
nature of the dark matter and for now we merely allow for the existence of 
{\it nonbaryonic} dark matter. Because regular clusters are so extended and 
practically nothing escapes from them during their evolution, it seems natural to 
suppose that they just reflect the mean composition of the whole 
Universe~\cite{kn:WNEF1993}. As a consequence, the {\it cluster} baryon fraction 
$f_B$ should equal the {\it cosmic} baryon fraction
\begin{equation}
f_B \simeq \frac{{\Omega}_B}{{\Omega}_M}~,   \label{NTee19aa3}  
\end{equation}
where ${\Omega}_B$ denotes the {\it baryonic} contribution to the cosmic density 
parameter. Once luminous matter in galaxies as well as possible 
baryonic dark matter are taken into account, eq. (\ref{NTee19aa2}) implies 
\begin{equation}
f_B \geq  0.17~.   \label{NTee19aa4}
\end{equation}
Combining eqs. (\ref{NTee19aa3}) and (\ref{NTee19aa4}) together, we find
\begin{equation}
{\Omega}_M \leq 5.9 \, {\Omega}_B~.   \label{NTee19aa5}
\end{equation}
We will use this result in Sect. 4.

Coming back to clusters of galaxies, another powerful tool to discover the 
presence of dark matter is {\it gravitational lensing}~\cite{kn:SEF1992}, namely 
the distortion of light rays when 
they pass close to a mass clump. Clusters act as {\it gravitational lenses}, which 
can magnify, distort and multiplay the images of background galaxies. A careful 
study of the properties of these images provides informations about the mass 
distribution inside the lens, and so ultimately on the existence of dark matter in 
clusters.

We begin by stressing that in a typical situation the distances background 
galaxy-cluster and cluster-observer are much larger than the size of the cluster 
itself. Accordingly, clusters behave as {\it thin lenses}, which means that 
lensing effects do not depend on the tridimensional mass distribution but rather 
on the {\it surface density} ${\Sigma}(R)$ on the sky. So, it is this quantity 
that can be derived from observations.

{\it Strong lensing}~\cite{kn:FM1994} -- This phenomenon concerns only {\it 
regular} clusters and is characterized by the 
existence of {\it giant arcs} inside the image of the cluster. Let us now discuss 
how these arcs arise and how they can be used to estimate the cluster mass.

In the first place, we recall that regular clusters possess spherical symmetry, 
namely {\it axial symmetry} about the line of sight to their centre. In such a 
situation, it can be shown that the {\it lens equation} takes the form
\begin{equation}
{\bf Y} \sim \left( 1 -  \frac{ m( < R)}{\pi  R^2  {\Sigma}_{cr}} \right) 
{\bf R}~,   \label{nt5}   
\end{equation} 
where $m( < R)$ denotes the mass inside the cylinder of radius $R$ about the line 
of sight
\begin{equation}
m( < R) \equiv 2 \pi \int_0^R ~ dR' \, R' \, {\Sigma}(R')~,   \label{nt6}   
\end{equation} 
while ${\Sigma}_{cr}$ stands for a reference value for the lens surface density 
which is completely fixed by the lens (cluster) and source (background galaxy) 
distances. In order to understand the meaning of eq. (\ref{nt5}), consider two 
planes orthogonal to the line of sight to the cluster centre: the {\it lens plane} 
and the {\it source plane}, whose meaning is clear from their names. Accordingly, 
${\bf Y}$ denotes a generic point $S$ in the source plane, whereas ${\bf R}$ 
represents the {\it image} $I$ of $S$ in the lens plane~\footnote{Both ${\bf Y}$ 
and ${\bf R}$ are measured from the point in which the line of sight intersecs the 
corresponding plane.}. Because of gravitational lensing, $S$ and $I$ are {\it 
misaligned} and $I$ is the point actually seen by the observer. It can also be 
shown that the {\it magnification} of the image is
\begin{equation}
\mu \sim \left( 1 - \frac{m( < R)}{\pi R^2 {\Sigma}_{cr}} \right)^{ - 1}~.   
\label{nt7}   
\end{equation} 
What is the image of that particular point ${\bf Y} = 0$ in which the line of sight 
intersects the source plane? By eq. (\ref{nt5}), we find that such an image is a 
whole circle -- the {\it Einstein ring} -- defined by the 
condition~\footnote{Obviously, the no-lensing solution ${\bf R} = 0$ is here 
unphysical and has to be discarded.}
\begin{equation}
m( < R) = \pi  R^2 {\Sigma}_{cr}~.   \label{nt8}   
\end{equation} 
Moreover -- owing to eq. (\ref{nt7}) -- this image has infinite magnification. 
Although such a divergence is merely an artifact of the geometric-optics 
approximation upon which eq. (\ref{nt7}) is based, the real magnification is 
nevertheless quite large. This circumstance qualifies the point ${\bf Y} = 0$ as a 
(degenerate) {\it caustic}. Hence, the image of a point source located just on the 
caustic is the Einstein ring. For an extended source close to the caustic, the 
situation is somewhat similar: the highly magnified image consists of {\it two 
elongated arcs}, which are sectors of the Einstein ring located on opposite sides. 
But even a tiny perturbation of the axial symmetry leads to a strong 
demagnification of one of these arcs, and so a {\it single arc} becomes the 
observational signature.

We are now in position to come back to the giant arcs seen in clusters of 
galaxies. It has been shown that their only physically consistent explanation is 
the above-discussed phenomenon, arising when a background galaxy lies occasionally 
close to the caustic, namely when it happens to be almost aligned with the cluster 
centre. Because it is easy to determine observationally both the radius of an arc 
and the distances of cluster and background galaxy~\footnote{In reality, 
background galaxies are too faint for their redshifts to be measured, but these 
quantities can be measured for their highly magnified images. Because 
gravitational lensing does not cause any frequency shift, the redshift of an arc 
yields the distance of the source through Hubble expansion law.}, such an 
interpretation leads -- by eq. (\ref{nt8}) -- to an estimate of the cluster mass 
enclosed by the corresponding Einstein ring.

Let us address the main limitations of this technique. First of all, the formation 
of giant arcs requires the existence of the caustic, which is a characteristic 
feature of the {\it strong lensing} regime. A necessary and usually sufficient 
condition for this to occur is that in at least one point of the lens the surface 
density exceeds ${\Sigma}_{cr}$ and it turns out that only regular clusters are 
dense enough to meet this constraint~\footnote{Equivalently -- once ${\Sigma}(R)$ 
and ${\Sigma}_{cr}$ are given -- eq. (\ref{nt8}) can be viewed as a condition for 
the formation of giant arcs. Only regular clusters fulfil such a condition.}. 
In addition, an almost perfect alignment observer-cluster-background galaxy is 
required, which is clearly a rather unlikely situation. Finally, only the cluster 
mass inside the Einstin ring can be estimated in this way.

{\it Weak lensing}~\cite{kn:BS2001} -- {\it Every} cluster produces {\it weakly 
distorted} images of all background galaxies that lie sufficiently close to its 
position on the sky. Because lensing compresses the image in one direction while 
stretching it in the orthogonal direction, the observed lensed images of backgroud 
galaxies are called {\it arclets}. Were galaxies perfectly round, the 
ellipticities of their arclets would tell us how strong the gravitational field is 
at every arclet position, from which the cluster mass can be derived. In reality, 
the unlensed image a generic galaxy has a nonvanishing ellipticity, which depends 
on the unknown galaxy orientation. So, the knowledge of the ellipticity of the 
corresponding arclet is useless. Still, suppose to contemplate {\it many} 
background galaxies at once. Then the {\it average} source ellipticity {\it 
vanishes}, since the individual ellipticities are manifestly randomly oriented and 
uncorrelated. Therefore, the average ellipticity of the corresponding arclets 
quantifies the weak lensing effect. What makes this technique effective is 
the existence of the so-calld {\it Tyson population} of faint blue 
galaxies with a surprisingly high surface number density, so that 
even a small patch of the sky -- over which ${\Sigma}(R)$ can be taken as constant 
-- is densely filled by them. Consequently, a map of the surface density 
${\Sigma}(R)$ on the sky can be derived from arclet observations. In spite of its 
conceptual simplicity, weak lensing is technically rather complex and will not be 
considered here any further.

We stress that -- at variance with the previous methods -- the mass determination 
based on gravitational lensing does {\it not} require any assumption about the 
dynamical state of the cluster. It is gratifying that even this strategy yields 
values of the mass-to-light ratio which generally {\it agree} with those derived 
by the virial analysis and $X$-ray studies. 

Specifically, the application of the above-discussed techniques to {\it regular} 
clusters gives values lying around the mean
\begin{equation}
{\Upsilon}_{RC} \simeq 210~.   \label{Znt8Z}   
\end{equation}
Because regular clusters contain a mix of spiral and elliptical 
galaxies~\footnote{They also contain lenticular galaxies, but this fact is 
unimportant for the present argument.} of similar luminosity, we expect the 
resulting ${\Upsilon}_{RC}$ to lie between ${\Upsilon}_S$ and ${\Upsilon}_E$. 
Recalling eqs. (\ref{ZZs35}) and (\ref{ZZs36AA}), this is just what eq. 
(\ref{Znt8Z}) tells us.

Clusters obviously contain the dark matter present in the cluster galaxies, but it 
might well happen that {\it additional} dark matter lurks in the intracluster 
space. However, an explicit analysis shows that the {\it total} galactic matter 
plus the hot gas {\it fully} account for the {\it total} cluster mass, thereby 
{\it ruling out} such a possibility~\cite{kn:BLD1995}.

\section{Cosmological evidence and nature of dark  matter}

Cosmology not only provides additional and dramatic evidence for the existence of 
a large amount of dark matter in the Universe, but also offers crucial 
informations about its physical nature. Cosmological implications for dark matter 
are nowadays rather well known in the particle physics community~\cite{kn:CPP} and 
are discussed in other talks at this conference. Moreover, a thorough account 
would be fairly complex. Because of these reasons, our analysis will be brief and 
rather schematic. Three specific items will be addressed, from which informations 
about the physical nature of dark matter will emerge.

\subsection{Primordial nucleosynthesis}

A crucial implication of the hot big bang cosmological model is that light 
elements -- like deuterium $D$, helium $He^3$, $He^4$ and lithium $Li^7$ -- must 
have formed during the first few minutes in the life of the Universe. Because the 
temperature monotonically decreases during the cosmic expansion, atomic nuclei can 
form when the energy of background photons becomes comparable to the nuclear 
binding energy. With the 
number of light neutrinos fixed to 3, the predicted light element abundances 
depend on a single free parameter, the cosmic baryon density ${\Omega}_B$. In 
fact, calculations show that an increase of ${\Omega}_B$ leads to slightly more 
$He^4$, but the resulting amounts of $D$ and $He^3$ drop dramatically. So, a 
comparison between the predicted and observed light element abundances 
unambiguously fixes ${\Omega}_B$~\cite{kn:NUCL}. Indeed, the agreement is achieved 
for ${\Omega}_B$ within a narrow range
\begin{equation}
{\Omega}_B \simeq 0.04 - 0.05~.   \label{nt11}   
\end{equation}
A remark is in order. No astrophysical process is known in which $D$ is produced, 
and so all the deuterium present in the Universe should be cosmological. 
Consequently, the comparison between theory and observation is particularly clean. 
In addition, local estimates of $D$ abundance are in good agreement with 
measurements in high-redshift clouds along the line of sight to a distant quasar. 

Regardless of big bang nucleosynthesis, an {\it independent} estimate of 
${\Omega}_B$ -- which turns out to agree with eq. (\ref{nt11}) -- arises from the 
features of high-redshift Lyman-$\alpha$ forest absorption lines of 
neutral hydrogen observed in the spectra of background quasars~\cite{kn:RAUCH}.

Before turning to a different argument, we stress that -- owing to eq. 
(\ref{nt11}) -- the upper bound (\ref{NTee19aa5}) becomes 
\begin{equation}
{\Omega}_M \leq  0.30~.   \label{NTee19aa5a}
\end{equation}

\subsection{Accelerated cosmic expansion}

Long ago, Hubble realized that informations about the geometry of the Universe can 
be obtained by observing {\it standard candles} -- astronomical objects of {\it 
known} absolute luminosity -- located at cosmological distances. Basically, the 
idea is as follows. Once we know the absolute luminosity of a source, the 
measurement of its apparent luminosity (radiative flux) yields the distance $D$. 
Moreover, it can be shown that $D$ depends -- in a known way -- on the source 
redshift $z$ through the parameters ${\Omega}_M$ and 
${\Omega}_{\Lambda}$~\footnote{Of course, for $z \ll 1$ one has $D \sim z$, which 
is the famous Hubble expansion law.}. In practice, a curve in the plot of apparent 
luminosity versus redshift -- the so-called {\it Hubble diagram} -- is labelled by 
the pair $({\Omega}_M,{\Omega}_{\Lambda})$. Suppose now that both apparent 
luminosity and redshift are measured for a sample of identical standard candles. 
Accordingly, a curve in the Hubble diagram gets singled out, and so a functional 
relationship between ${\Omega}_M$ and ${\Omega}_{\Lambda}$ emerges.

A few years ago, this strategy has been applied to a sample of distant Type-Ia 
supernovae, which are believed to be good standard candles~\cite{kn:RP}. A 
best-fit to the data yields
\begin{equation}
{\Omega}_{\Lambda} \simeq 1.33 \, {\Omega}_M + 0.33~.   \label{nt11a}   
\end{equation}
Because eq. (\ref{nt11a}) entails ${\rho}_{\Lambda} > 0$ and $\rho_M < 2 \, 
{\rho}_{\Lambda}$, we are forced to conclude that the present cosmic expansion is 
{\it accelerated}.

Observe that -- by eq. (\ref{nt11a}) -- the upper bound (\ref{NTee19aa5a}) yields
\begin{equation}
{\Omega}_{\Lambda} \leq  0.73~.   \label{zNTee19aa5a}
\end{equation}

\subsection{Cosmic Microwave Background}

Another fundamental implication of big bang cosmology is the existence of the {\it 
Cosmic Microwave Background} (CMB), which is just primordial light redshifted by 
the cosmic expansion. Besides providing a wonderful confirmation of the big bang 
paradigm, the CMB yields a wealth of informations about dark matter.

Much in the same way as nuclei form once the temperature drops below the nuclear 
binding energy, atoms come into existence when the temperature becomes comparable 
to the atomic binding energy. When this process -- named {\it recombination} -- 
takes place, matter becomes neutral and decouples from radiation, which 
accordingly streams freely throughout the Universe. As a result, the CMB brings to 
us a snapshot of the Universe just at recombination, namely at redshift $z_{rec} 
\simeq 1100$ (corresponding to $t_{rec} \simeq 3 \cdot 10^5$ years after the big 
bang). 

Soon after its discovery in 1965, it was established that the CMB has a black body 
spectrum peaked at $T_0 \simeq 2.72 \, K$. Early analyses showed that the CMB is 
highly isotropic, once our peculiar motion -- producing a {\it systematic 
anisotropy} with ${\Delta} T/T_0 \sim 10^{- 3}$ -- is corrected for. 
However, in 1992 the COBE mission discovered {\it anisotropies} in the CMB 
spectrum, corresponding to temperature {\it fluctuations} ${\Delta} T/T_0 \sim 
10^{- 5}$ on the angular scale of $7^0$. Remarkable progress has been made in the 
last few years, with the BOOMERANG, MAXIMA and DASI missions detecting similar 
temperature fluctuations down to $1^0$~\cite{kn:BMD}. Quite recently, the WMAP 
mission has succeeded in discovering temperature fluctuations on the angular scale 
of $0.2^0$~\cite{kn:WMAP2}.

What is the physical meaning of the CMB fluctuations? Because the 
post-recombination Universe is essentially transparent to the CMB photons, those 
which we detect now had their last interaction with matter on a virtual sphere -- 
centered at our position -- named {\it last scattering surface} (LSS). At the time 
of recombination, a generic point of the LSS had an horizon of size $d_h \simeq c 
\, t_{rec}$, which we see today under an angle ${\theta}_1$. Therefore, only 
events lying within ${\theta}_1$ -- about a given direction in the sky -- were 
causally connected at recombination. As a consequence, only CMB fluctuations  
on angular scales $\theta < {\theta}_1$ yield informations about physical 
processes occurring during recombination. However, recombination was not an 
instantaneous process and this fact implies that the LSS is a shell of 
{\it finite} thickness, corresponding to an observed angular scale ${\theta}_2$ 
(obviously $\theta_2 < \theta_1$). As a result, CMB fluctuations get smeared out 
over angular scales $\theta < {\theta}_2$. Thus, we conclude that recombination 
physics shows up in CMB fluctuations on angular scales in the range $\theta_2 < 
\theta < \theta_1$. These fluctuations are the imprint on the CMB of {\it acoustic 
oscillations} in the matter-radiation fluid just before decoupling, with gravity 
providing the driving force while radiation pressure causes the restoring 
one~\cite{kn:WSS1997}.

A quantitative description of the CMB fluctuations~\cite{kn:PAD2002} emerges from 
a statistical 
treatment based on the harmonic analysis of ${\Delta} T/ T_0$. Were measurements 
be performed on a plane, ${\Delta} T/ T_0$ would depend on $x, y$ and we would 
represent ${\Delta} T/ T_0 (x,y)$ as a Fourier series. But in reality 
${\Delta} T/ T_0$ is measured on the celestial sphere and so it depends on 
$\theta, \varphi$. Accordingly, the following multipole expansion naturally arises
\begin{equation}
\frac{\Delta T}{T_0} (\theta, \varphi) = \sum_{l=0}^{\infty} 
\sum_{m=-l}^{l} a_{lm} Y_{lm} (\theta, \varphi)~,    \label{ntc11}   
\end{equation}
where $Y_{lm} (\theta, \varphi)$ are spherical harmonics and the coefficients 
$a_{lm}$ are {\it gaussian random variables} defined by
\begin{equation}
\langle  a_{lm} \rangle =0~,   \label{ntc12}   
\end{equation}
\begin{equation}
\langle a_{lm} a^*_{l'm'} \rangle = c_l \delta_{ll'} \delta_{mm'}~,  \label{ntc13}
\end{equation}
with $\langle \cdots \rangle$ representing the average over the whole sky. 
Denoting by $\alpha$ the angle between two arbitrary directions ${\hat n} \equiv 
(\theta , \varphi)$ and ${\hat n}' \equiv (\theta' , \varphi')$  ($\cos \alpha = 
{\hat n} \cdot {\hat n}'$), the CMB {\it autocorrelation function} is 
\begin{equation}
C (\alpha) \equiv \left\langle \frac{\Delta T}{T} (\theta, \varphi) \, 
\frac{\Delta T}{T} (\theta', \varphi') \right\rangle~.  \label{ntc14}   
\end{equation}
It can be shown that the autocorrelation function can be represented in terms of 
the {\it multipole moments} $c_l$ as
\begin{equation}
C(\alpha) = \frac{1}{4 \pi} \sum_{l=0}^{\infty} (2l +1) c_l P_l (\cos \theta)~,  
\label{ntc15}   
\end{equation}
where $P_l(\cos \theta)$ are Lagendre polinomials. It turns out that each term in 
eq. (\ref{ntc15}) corresponds to a well-defined angular scale, given by
\begin{equation}
\theta \sim \frac{180^0}{l}~.   \label{ntc16}   
\end{equation}
Hence, fluctuations on {\it small} angular scales correspond to {\it large} 
multipole orders (and vice-versa). Consider now the CMB {\it power spectrum}, 
namely the graph of $l (l + 1) c_l$ versus $l$, and denote by $l_1$ and $l_2$ the 
multipole orders corresponding -- by eq. (\ref{ntc16}) -- to $\theta_1$ and 
$\theta_2$ (respectively). Then the above-discussed acoustic oscillations show up 
as {\it acoustic peaks} in the CMB power spectrum within the interval 
$l_1 < l < l_2$. It is precisely these peaks that tell us much about dark matter.

Specifically, the {\it position} of the first peak is controlled by $\theta_1$. 
Being an angle, $\theta_1$ is very sensitive to the geometry of the Universe, that 
is to say to $\Omega$. So, the actual position of the first peak yields the 
specific value of $\Omega$. The result is~\cite{kn:BMD}~\cite{kn:WMAP}
\begin{equation}
\Omega \simeq 1~.   \label{ntc17}   
\end{equation}
Thanks to eq. (\ref{Zco4c3}), eq. (\ref{ntc17}) entails that $\Omega_M$ and 
${\Omega}_{\Lambda}$ have to meet the constraint
\begin{equation}
\Omega_M + {\Omega}_{\Lambda} \simeq 1~.   \label{RRntc17}   
\end{equation}
Moreover, it can be shown that the {\it ratio} of the {\it heights} of the first 
to the second peak gives~\cite{kn:BMD}~\cite{kn:WMAP} 
\begin{equation}
{\Omega}_B \simeq 0.045~.  \label{ntc19}   
\end{equation}

\subsection{Discussion}

Let us now try to combine the various pieces of information obtained above into a 
coherent cosmological setting.

Perhaps, the most dramatic result is expressed by eq. (\ref{ntc17}). For, it 
implies that the Universe is spatially {\it flat} -- namely {\it euclidean} -- in 
agreement with the natural expectation based on cosmic inflation~\cite{kn:LL}. 

Another remarkable fact is that the value of $\Omega_B$ -- as given by eq. 
(\ref{ntc19}) -- nicely fits within the range (\ref{nt11}). Hence, we see that 
cosmology provides a solid prediction about the total amount of baryons in the 
Universe. 

Combine next eq. (\ref{nt11a}) with eq. (\ref{RRntc17}). Then it follows 
\begin{equation}
{\Omega}_M \simeq 0.29~,   \label{ntc18}   
\end{equation}
\begin{equation}
{\Omega}_{\Lambda} \simeq 0.71~.   \label{ntc18a2}   
\end{equation}
In the first place, observe that eqs. (\ref{ntc18}) and (\ref{ntc18a2}) are 
consistent with the upper bounds (\ref{NTee19aa5a}) and (\ref{zNTee19aa5a}) 
(respectively). As far as $\Omega_M$ is concerned, several other independent 
estimates exist, which are based mainly on galaxy surveys and observations of 
large-scale structure~\cite{kn:SS}. They cluster around eq. (\ref{ntc18}), 
however with considerable scatter. Once allowance for experimental errors is made, 
the resulting individual values of $\Omega_M$ tend to agree with eq. (\ref{ntc18}) 
and the upper bound (\ref{NTee19aa5a}) is (marginally) satisfied.

So, we see that all the various contributions to the $\Omega$ parameter turn 
out to fit within a {\it consistent} cosmological scenario, the so-called {\it 
concordance cosmology}~\cite{kn:CC}.

\subsection{Nature of dark matter}

Finally, we briefly address the nature of dark matter as implied by the above 
analysis.

{\it Baryonic dark matter} -- We estimated in Sect. 2 the contribution 
${\Omega}_*$ to the $\Omega$ parameter from {\it luminous} matter -- see eq. 
(\ref{mo3}) -- and we found that ${\Omega}_*$ is smaller than ${\Omega}_B$ -- 
as fixed by eq. (\ref{ntc19}) -- by a factor of 10. Hence, about $90 \, {\%}$ of the 
baryons do not emit photons in the {\it optical} band. What about the rest of the 
baryons? A recent inventory~\cite{kn:FHP1998} of the baryonic content of the 
Universe -- obtained by combining available observational data with the whole body 
of theoretical knowledge -- shows that the observed baryon budget today is 
dominated by $X$-ray emitting hot gas in groups and clusters of galaxies and 
accounts for almost $40 \, {\%}$ of the ${\Omega}_B$ value dictated by eq. 
(\ref{ntc19}). Thus, nearly {\it half} of the existing baryons are invisible and 
make up {\it baryonic dark matter}. We do not discuss this topic any further and 
simply refer to our review paper~\cite{kn:io}.

{\it Nonbaryonic dark matter} -- As the reader will certainly have noticed, 
$\Omega_M$ turns out to largely exceed $\Omega_B$. Therefore, 
dark matter -- as clumped in galaxies -- is {\it dominated} by elementary particles 
carrying no baryon number, the so-called WIMPs (weakly interacting massive 
particles)~\footnote{See the contribution of R. Bernabei to these proceedings.}. In 
this way, a deep connection emerges between cosmology and particle 
physics~\footnote{See the contribution of A. Masiero to these proceedings.}. 
Besides, it should be emphasized that the existence of nonbaryonic dark matter is 
also required by an {\it independent} argument, namely {\it galaxy formation}. 
Indeed, it would be impossible to explain the existence of structure in the present 
Universe with baryons alone~\cite{kn:LONGAIR}. 

{\it Dark energy} -- This is the stuff responsible for the {\it accelerated} 
cosmic expansion and described by ${\Omega}_{\Lambda}$. In Sect. 2 we assumed that 
${\Omega}_{\Lambda}$ arises from the vacuum energy represented by a cosmological 
constant, but this view is too restricted. In fact, dark energy can be anything 
that emits no light, has {\it negative} pressure and does not significantly 
cluster on the Megaparsec scale. Incidentally, the latter fact can be viewed as a 
natural consequence of the negative pressure~\footnote{See the contribution of S. 
Matarrese to these proceedings.}. 

Thus, finding the dark baryons, detecting WIMPs and discovering the specific 
properties of the dark energy are crucial observational challanges for contemporary 
cosmology.

\section{Acknowledgements}

We thank professor Milla Baldo Ceolin for her kind invitation to talk at 
this splendid conference.


\begin{thebibliography}{99}

\bibitem{kn:PEACOCK} S. Weinberg, {\it Gravitation and Cosmology} (Wiley, New 
York, 1972). P. J. E. Peebles, {\it Principles of Physical Cosmology} 
(Princeton University Press, Princeton, 1993). J. A. Peacock, {\it Cosmological 
Physics} (Cambridge University Press, Cambridge, 1999). P. Coles and F. Lucchin, 
{\it Cosmology} (Wiley, New York, 2002). J. V. Narlikar, {\it An Introduction to 
Cosmology} (Cambridge University Press, Cambridge, 2002).

\bibitem{kn:FI1986} J. E. Felten and R. Isaacson, {\it Rev. Mod. Phys.} {\bf 58} 
(1986) 689.

\bibitem{kn:LONGAIR} M. S. Longair, {\it Galaxy Formation} (Springer, Berlin, 1998).

\bibitem{kn:FHP1998} M. Fukugita, C. J. Hogan and P. J. E. Peebles, {\it 
Astrophys. J.} {\bf 503} (1998) 518.

\bibitem{kn:BT1987} J. J. Binney and S. Tremaine, {\it Galactic Dynamics} 
(Princeton University Press, Princeton, 1987).

\bibitem{kn:RUBIN} K. C. Freeman, {\it Astrophys. J.} {\bf 160} (1970) 811. V. C. 
Rubin, D. Burstein, W. K. Ford and N. Thonnard, {\it Astrophys. J.} {\bf 289} 
(1985) 81. D. Burstein and V. C. Rubin, {\it Astrophys. J.} {\bf 297} (1985) 423. 
Y. Sofue and V. C. Rubin, {\it astro-ph/0010594} (2000). S. M. Kent, {\it Astron. 
J.} {\bf 91} (1986) 1301.

\bibitem{kn:vbbs1985} A. Bosma, {\it Astron. J.} {\bf 86} (1981) 1825. T. S. van 
Albada, J. N. Bahcall, K. Begeman and R. Sancisi, {\it Astrophys. J.} {\bf 295} 
(1985) 305.

\bibitem{kn:BT1981} J. N. Bahcall and S. Tremaine, {\it Astrophys. J.} {\bf 244} 
(1981) 805.

\bibitem{KN:Zw1994} D. Zaritsky and S. D. M. White, {\it Astrophys. J.} {\bf 435} 
(1994) 599.

\bibitem{kn:BLD1995} N. A. Bahcall, L. M. Lubin and V. Dorman, {\it Astrophys. J.} 
{\bf 447} (1995) L81.

\bibitem{kn:SAGLIA} R. P. Saglia, G. Bertin and M. Stiavelli, {\it Astrophys. J.} 
{\bf 384} (1992) L433.

\bibitem{kn:GERHARD} O. Gerhard {\it et al.}, {\it Astron. J.} {\bf 121} (2001) 
1936. M. Capaccioli, N. R. Napolitano and M. Arnaboldi, {\it astro-ph/0211323} 
(2002).

\bibitem{TREU} T. Treu and L. V. E. Koopmans, {\it astro-ph/0202342} (2002) .

\bibitem{kn:FJT1985} W. Forman, C. Jones and W. Tucker, {\it Astrophys. J.} 
{\bf 293} (1985) 102. C. R. Canizares, G. Fabbiano and G. Trinchieri, {\it 
Astrophys. J.} {\bf 312} (1987) 503. G. Fabbiano, {\it Annu. Rev. Astron. 
Astrophys.} {\bf 27} (1989) 87.

\bibitem{kn:OPC2003} E. O'Sullivan, T. J. Ponman and R. S. Collins, {\it 
astro-ph/030115 } (2003).

\bibitem{kn:SLE} R. E. Griffiths, S. Casertano, M. Im and K. U. Ratnatunga, 
{\it Mon. Not. R. Astron. Soc.} {\bf 282} (1996) 1159. G. Wilson, N. Kaiser, G. 
Luppino and L. L. Cowie, {\it astro-ph/0008504} (2001).

\bibitem{kn:RPK1972} H. J. Rood, T. L. Page and E. C. Kinter, {\it Astrophys. J.} 
{\bf 175} (1972) 627.

\bibitem{kn:CARLBERG1995} R. Carlberg {\it et al.}, {\it Astrophys. J.} {\bf 462} 
(1996) 32.

\bibitem{kn:WNEF1993} S. D. M. White, J. F. Navarro, A. E. Evrard and C. S. Frenk, 
{\it Nature} {\bf 366} (1993) 429.

\bibitem{kn:SARAZIN} C. L. Sarazin, {\it Rev. Mod. Phys.}  {\bf 58} (1986) 1.

\bibitem{kn:SEF1992} P. Schneider, J. Ehlers and E. Falco, {\it Gravitational 
Lenses} (Springer, Berlin, 1992).

\bibitem{kn:FM1994} B. Fort and Y. Mellier, {\it Astron. Astrophys. Rev.} {\bf 5} 
(1994) 239.

\bibitem{kn:BS2001} M. Bartelmann and P. Schneider, {\it Phys. Rep.} {\bf 340} 
(2001) 291.

\bibitem{kn:CPP} G. Steigman, {\it Ann. Rev. Nucl. Part. Sci.} {\bf 29} (1979) 
313. J. R. Primack, D. Seckel and B. Sadoulet, {\it Ann. Rev. Nucl. Part. Sci.} 
{\bf 38} (1988) 751. E. W. Kolb and M. S. Turner, {\it The Early Universe} 
(Addison-Wesley, New York, 1990). L. Bergstrom and A. Goobar, {\it Cosmology and 
Particle Physics} (Wiley, New York, 1999). M. S. Turner and J. A. Tyson, {\it Rev. 
Mod. Phys.} {\bf 71} (1999) S145.

\bibitem{kn:NUCL} D. N. Schramm and M. S. Turner, {\it Rev. Mod. Phys.} {\bf 70} 
(1998) 303. K. A. Olive, G. Steigman and T. P. Walker, {\it Phys. Rep.} {\bf 333} 
(2000) 389. D. Tytler, J. M. O'Meara, N. Suzuki and D. Lubin, {\it Phys. Rep.} 
{\bf 333} (2000) 409. A. D. Dolgov, {\it Nucl. Phys. B (Proc. Suppl.)} {\bf 110} 
(2002) 137.

\bibitem{kn:RAUCH} M. Rauch {\it et al.}, {\it Astrophys. J.} {\bf 489} (1998) 1.

\bibitem{kn:RP} A. Riess {\it et al.}, {\it Astron. J.} {\bf 116} (1998) 1009. S. 
Perlmutter {\it et al.}, {\it Astrophys. J.} {\bf 517} (1999) 565.

\bibitem{kn:LL} A. R. Liddle and D. H. Lyth, {\it Cosmological Inflation and 
Large-Scale Structure} (Cambridge University Press, Cambridge, 2000). A. Riotto, 
{\it hep-ph/0210162} (2002).

\bibitem{kn:SS} S. Schindler, {\it astro-ph/0107028} (2001).

\bibitem{kn:io} M. Roncadelli, {\it astro-ph/0301537} (2003).

\bibitem{kn:BMD} A. Balbi {\it et al.}, {\it Astrophys. J.} {\bf 545} (2000) L1. 
P. De Bernardis {\it et al.}, {\it Astrophys. J.} {\bf 564} (2002) 559. C. Pryke 
{\it et al.}, {\it Astrophys. J.} {\bf 568} (2002) 46.

\bibitem{kn:WMAP2} C. L. Bennet {\it et al.}, {\it astro-ph/0301158} (2003).


\bibitem{kn:WSS1997} W. Hu, N. Sugiyama and J. Silk, {\it Nature} {\bf 386} (1997) 
37.

\bibitem{kn:PAD2002} W. Hu and S. Dodelson, {\it astro-ph/0110414} (2001). T. 
Padmanabhan, {\it Theoretical Astrophysics, vol. 3} (Cambridge University Press, 
Cambridge, 2002). 

\bibitem{kn:WMAP} D. N. Spergel {\it et al.}, {\it astro-ph/0302209} (2003).


\bibitem{kn:CC} J. P. Ostriker and P. J. Steinhardt, {\it Nature} {\bf 377} (1995) 
600. N. A. Bahcall, J. P. Ostriker, S. Perlmutter and P. J. Steinhardt, {\it 
Science} {\bf 284} (1999) 1487. L. Wang, R. R. Caldwell, J. P. Ostriker and P. J. 
Steinhardt, {\it Astrophys. J.} {\bf 530} (2000) 17.


\end{thebibliography}
\end{document}